\documentclass[preprint,12pt,aps,prd]{revtex4-1}
\usepackage{amsmath}
\usepackage{graphicx}
\usepackage{amssymb}
\usepackage{hyperref}
\usepackage{float}
\usepackage[T1]{fontenc}
\usepackage{tabularx,caption}
\begin{document}

\def\be{\begin{equation}}
\def\ee{\end{equation}}
\def\ba{\begin{array}}
\def\ea{\end{array}}
\def\bc{\begin{center}}
\def\ec{\end{center}}
\newcommand{\ra}{\rangle}
\newcommand{\la}{\langle}
\newcommand{\eq}{\begin{eqnarray}}
\newcommand{\en}{\end{eqnarray}}
\newcommand{\bfq}{{\bf q}_{\perp}}
\newcommand{\bfql}{{\bf q}_{1\perp}}
\newcommand{\bfqr}{{\bf q}_{2\perp}}
\newcommand{\bfk}{{\bf k}_{\perp}}
\newcommand{\bfkpr}{{\bf k}_{\perp}^\prime}
\newcommand{\bfki}{{\bf k}_{\perp i}}
\newcommand{\bfb}{{\bf b}_{\perp}}
\newcommand{\bfbi}{{\bf b}_{\perp i}}
\newcommand{\bfe}{{\bf e}_{\perp}}
\newcommand{\bfP}{{\bf P}_{\perp}}
\newcommand{\bfPpr}{{\bf P}_{\perp}^\prime}
\newcommand{\bfp}{{\bf p}_{\perp}}
\newcommand{\bfn}{{\bf n}}
\newcommand{\bfS}{{\bf S}_{\perp}}
\newcommand{\bfppr}{{\bf p}_{\perp}^\prime}
\newcommand{\bfQ}{{\bf Q}_{\perp}}
\newcommand{\bfQpr}{{\bf Q}_{\perp}^\prime}
\newcommand{\bfD}{{\bf \Delta}_{\perp}}
\newcommand{\bfr}{{\bf r}_{\perp}}
\title{Transverse momentum dependent parton distributions of pion in the light-front holographic model}
\author{Navdeep Kaur and Harleen Dahiya}
\affiliation{Department of Physics, Dr. B. R. Ambedkar National Institute of Technology, Jalandhar-144011, India}
\begin{abstract}
Using the light-front holographic model, we study the transverse momentum dependent parton distributions (TMDs) for the case of pion. At leading twist, the  unpolarized parton distribution function $ f_{1\pi}(x,\bfk^{2}) $  and  the Boer-Mulders function $ h_{1\pi}^{\bot}(x,\bfk^{2}) $ are obtained for  pion. We calculate both the functions using the light-front holographic model with spin improved wave function and compare the predicted results with available results of other models. In order to provide inputs in predicting  future experimental data, a LO evolution is performed from model scale to experimental scale for the case of unpolarized parton distribution function $ f_{1\pi}(x,\bfk^{2}) $.
\end{abstract}
\maketitle
\section{Introduction}
Understanding the hadronic structure in terms of its constituents  has become the major goal of high energy physics during the last few decades. A lot of experimental as well as theoretical efforts are continuously being made to explore the three-dimensional (3D) picture of hadrons in configuration and momentum  space. A one-dimensional picture of hadron is provided by the parton distribution functions (PDFs)  which describe the probability of finding a parton (quark or gluon) with longitudinal momentum fraction inside the hadron in forward scattering region. By extending the PDFs from forward scattering region to off-forward scattering region, one can obtain the generalized parton distributions (GPDs). The GPDs not only provide the information about the longitudinal momentum fraction of parton but also give the longitudinal momentum transfer between the initial and final state of hadron \cite{burk2000}. The various scattering process like deep virtual Compton scattering (DVCS), deep virtual meson production (DVMP) provide detailed information on the GPDs \cite{Ji:1997, Ji2:1997, Radyushkin:1997, Ji:1998, Ji2:1998, Blmlein:2000, Goeke:2001, Diehl:2003}. Further, the study of transverse momentum dependent parton distribution function (TMDs) provide the 3D picture of the hadrons and related information about the correlation between the spin of parent hadron and the intrinsic transverse momentum of its constituent partons \cite{Tangerman95, Kotzinian95, Mulders97, Mulders98, Mulders07}.
In recent years, several experiments have been running and are being planned for future to gather such information. 

The TMDs of hadrons can be measured experimentally in semi-inclusive deep inelastic scattering (SIDIS) \cite{Mulders07, Ji:2005} and Drell-Yan (DY) processes \cite{Drell:1970, Christenson:1970, Collins:2002, Quintans:2011, Baranov:2014}. TMDs allow us to access the partonic configuration of hadrons with  longitudinal momentum fraction $ x $ and transverse momentum $ \bfk $. Apart from $ x $ and $ \bfk $, TMDs also depend on the Wilson line (gauge link). The presence of Wilson line in TMD correlator is necessary to attain color gauge invariance and to obtain non-zero T-odd TMDs (Sivers and Boer-Mulders functions). Wilson lines are process dependent so their presence in TMD correlator makes T-odd TMDs process dependent. In the case of SIDIS process  final state interaction (FSI) and in the case of DY process the initial state interaction  between struck quark and spectator via one gluon exchange  are necessary for obtaining non-vanishing T-odd TMDs \cite{Brodsky:2002, Ji:2002, Brodsky:2002(2), Boer:2003, Boer:2003(2), Belitsky:2003}.
 
Pion being the simplest QCD bound state of quark and antiquark among the hadrons, has attracted lot of attention and also they have the advantage of being probed experimentally. In DY process, the structure of pion can be probed by hitting them on nuclear target \cite{McGaughey:1999, Reimer:2007, Chang:2013, Peng:2014}. In past years, lot of theoretical work has been done to explore the pion structure including the studies on pion PDFs in the light-front constituent model, chiral constituent quark model  \cite{Altarelli:1996, Chang:2014, Shi:2018}, Nambu-Jone-Lasino model and  using Dyson–Schwinger equation \cite{Nguyen:2011, Chang:2015} and Bethe-Salpeter wave functions. GPDs of pion have also been studied extensively  in various other models \cite{Polyakov:1999, Dalley:2001, Davidson:2002, Dalley:2003, Broniowski:2003, Bromel:2008, Frederico:2009, Dorokhov:2011, Mezrag:2014, Thomas:2015}. Recently pion TMDs have been calculated in light-front constituent approach \cite{Barbara:2014, Lorce:2016}, spectator model \cite{Lu:2004, Meissner:2008, Gamberg:2010}, in NJL model \cite{Noguera} and in AdS/QCD approach \cite{Bacchetta:2017, Kaur:2018}. 

To study 3D structure of hadrons theoretically, hadronic wave functions in terms of quark and gluons degree of freedom are important. There have been several attempts in different models to obtain the hadronic wave function, the most recent being the AdS/QCD correspondence in light-front formalism.  By holographic mapping between the Hamiltonian formulation for quantized QCD and specific properties for hadron in five-dimensional AdS-space one can obtain the hadronic wave function in light-front holographic model. The solution of the holographic Schr\"odinger equation for meson within the semiclassical approximation of light-front QCD yields the valence meson wave function for massless quarks \cite{Brodsky2005, Brodsky2006, Brodsky2009, Brodsky2014}. For non-vanishing quark masses, there is a  need to go beyond the semiclassical approximation \cite{Brodsky15}. This has been done by including a perturbation mass term to the effective potential of the holographic Schr\"odinger equation \cite{Brodsky:2009, Vega:2009, Vega:2009(2), Rohit:2015}. Within the semiclassical approximation of light-front QCD, the dynamical spin effects are usually neglected. For pion, such effects have been taken into account in recent work \cite{Ahmady:2017, Ahmady:2018}. 

In order to study the spin structure of pion, we have used  the holographic light-front wave function with dynamical spin from the light-front holographic model \cite{Ahmady:2018} which includes the higher helicity (parallel spin of partons) components besides the usual lower helicity (anti-parallel spin of partons) components.  The higher helicity components may have the possibility to  provide non-leading contribution in the light-front wave-function. The main motivation to choose this model is the presence of spin wave function in holographic light-front wave-functions  which is not usually present in the original AdS/QCD model but is necessary to obtain the Boer-Mulders function of pion. This model gives a simultaneous successful description of wide range of experiment data including the decay constant, the charge radius and EM form factor of pion \cite{Ahmady:2018}.

In the present work, we predict the results for pion TMDs in light-front holographic with spin-improved wave function.  At leading twist, the pion transverse dependent quark-quark correlator consists of two TMDs, the unpolarized parton distribution function $ f_{1 \pi}^{q}(x,\bfk^{2}) $ and the Boer-Mulders function $ h_{1 \pi}^{\bot}(x,\bfk^{2}) $. The TMD $ f_{1 \pi}^{q}(x,\bfk^{2}) $ has been obtained from the lowest order correlation function. We have compared our result with the previous results from light-front constituent model \cite{Barbara:2014} and soft wall AdS/QCD \cite{Kaur:2018}. To obtain the function $ h_{1 \pi}^{\bot}(x,\bfk^{2}) $, we have used the higher order correlation function to consider the contribution from the one-gluon exchange between quark and antiquark. The obtained $ h_{1 \pi}^{\bot}(x,\bfk^{2}) $ function has also been compared with the results of light-front constituent model. In context of experimental observations, the evolution of TMDs are very crucial for future comparisons. However, as the exact evolution equation for the Boer-Mulders functions is still under study, we present here the case for only the  unpolarized parton distribution function $ f_{1 \pi}^{q}(x,\bfk^{2}) $. We show how it evolves from initial model scale to higher energy scale using the evolution equation.
 
The outline of this paper is as follows. In Section II, we discuss the light-front amplitudes in terms of pion light-front wave functions(LFWFs) with different orbital angular momentum components. In Section III, the unpolarized parton distribution function of pion is predicted at model scale. The Boer-Mulders function of pion is given in Section IV. In Section V we discuss the TMD evolution of the unpolarized parton distribution function. We summarize and conclude our results in Section VI.

\section{Light-front amplitudes of the pion in light-front holographic model}
For the description of a hadron in the light-front dynamics, the classification of independent wave-function amplitudes with a particular parton combination is necessary. With the information  about the parton content of a hadron, one can obtain the  light-front amplitudes for a hadron. The light-front amplitudes of a pion with helicity $\Lambda $ can be written in terms of total parton helicity $ \lambda $ and total orbital angular momentum of partons $ L_{z} $. With the lowest quark-antiquark Fock state component, the total parton helicity of pion can be either $0$ or $1$. Correspondingly, according to law of conservation of angular momentum, $ \Lambda = \lambda + L_{z} $. Therefore,  two light-front wave function amplitudes with total quark orbital momentum $ L_{z}=0 $ and $ \mid L_{z} \mid = 1 $ exist for a pion with valence $\vert q \bar{q}\rangle $ Fock state. The light-front wave function for the quark-antiquark Fock state of a positively charged pion with momentum $ P $ can be written as the superposition of these light-front amplitudes as follows 
\eq
|\pi^+(P)\ra &=& \ |\pi^+(P)\ra_{L_z=0} +
|\pi^+(P)\ra_{|L_z|=1} \,,\nonumber\\
|\pi^+(P)\ra_{L_z=0}  &=&
\int \frac{d^2\bfk}{16\pi^3} \, \frac{dx}{\sqrt{6 x (1-x)}} \,
\Psi_\pi^{(0)}(x, \bfk) \, \sum_{a=1}^{3} \,
\Big[b^{\dagger a}_{u\uparrow}(1) d^{\dagger a}_{d\downarrow}(2)
-    b^{\dagger a}_{u\downarrow}(1)
     d^{\dagger a}_{d\uparrow}(2)\Big] \, |0\ra, \nonumber\\
|\pi^+(P)\ra_{|L_z|=1}   &=&
\int \frac{d^2\bfk}{16\pi^3} \, \frac{dx}{\sqrt{2 x (1-x)}} \,
\Psi_\pi^{(1)}(x, \bfk) \nonumber\\
&\times&  \, \Big(\frac{k_x + ik_y}{\sqrt{3}} \, \sum_{a=1}^{3} \,
     b^{\dagger a}_{u\uparrow}(1)
     d^{\dagger a}_{d\uparrow}(2) \, |0\ra\,
+  \frac{k_x - ik_y}{\sqrt{3}} \, \sum_{a=1}^{3} \,
    b^{\dagger a}_{u\downarrow}(1)
     d^{\dagger a}_{d\downarrow}(2)  \, |0\ra\,\Big),
\label{light-front-amplitude}
\en
where $1 = (x,\bfk)$ and $2 = ((1-x),-\bfk)$ are the coordinates of quark and antiquark respectively. Here $ \Psi_\pi^{(0)}(x, \bfk) $ and $\Psi_\pi^{(1)}(x, \bfk) $ are the pion LFWFs  for total quark orbital momentum $ L_{z}=0 $ and $ \mid L_{z} \mid = 1 $ respectively. The creation (annihilation) operators of $u$-quark and $\bar d$-quark with color $a$ are  $b^{\dagger a}_\lambda(b^a_\lambda)$ and $d^{\dagger a}_\lambda(d^a_\lambda)$ respectively.

The pion state with leading $ \vert q\bar{q}\rangle $ Fock state component can also be written as \cite{Barbara:2010}
\eq
|\pi^+(P)\ra  &=&
\sum_{\lambda_i} \,\int \frac{d^2\bfk}{16\pi^3} \, \frac{dx}{\sqrt{x (1-x)}} \,
\Psi_\pi^{L_z}(x_{i},\textbf{k}_{i\bot}; \lambda_i) \,\frac{\delta_{ij}}{\sqrt{3}}\,
b^{\dagger a}_{u \lambda_1}(1) d^{\dagger a}_{d \lambda_2}(2) |0\ra,
\label{pion-state}
\en
where $ x_i $ and $ \textbf{k}_{i\bot} $ are the longitudinal momentum fraction and transverse momentum of constituents (quark and antiquark) of pion, respectively. The LFWF  $ \Psi_\pi^{L_z}(x_i,\textbf{k}_{i\bot};\lambda_i) $ in  Eq. \eqref{pion-state} can be constructed as the product of a momentum-dependent wave function and a spin-dependent wave function. We have 
\eq
\Psi_\pi^{L_z}(x_i,\textbf{k}_{i\bot};\lambda_i)&=& \psi_\pi (x,\bfk)\, \phi_\pi(\lambda_{1},\lambda_{2}). 
\en
For the momentum-dependent part of wave function, we choose the light front holographic wave function of meson as
\eq
 \psi_{\pi} (x,\bfk) = \frac{4 \pi N_{0}}{\kappa \sqrt{x (1-x)}}   \exp{ \left[ -\frac{\bfk^2 + m^2}{2 \kappa^2 x(1-x)} \right]}. 
 \label{LFHM}
\en
Here $ N_{0} $ is the normalization constant, $\kappa$ is the scale parameter and $m$ is the quark mass. The spin dependent part of wave function is defined as 
\eq
	\phi_\pi(\lambda_{1},\lambda_{2})= \frac{\bar{u}_{\lambda_{1}}(x,\bfk)}{\sqrt{\bar{x}}} \left[A \frac{M_P}{2P^+} \gamma^+ \gamma^5 + B  \gamma^5 \right] \frac{v_{\lambda_{2}}(x,\bfk)}{\sqrt{x}}. 
\label{spin-structure} 
\en
Here $ A $ and $ B $ are constants whose values are constrained by the decay constant and charged radii of pion. In the present work we use $ A=1 $ and $ B = 1 $ following Ref. \cite{Ahmady:2017} as the results of decay constant and radius of charged pion obtained in  this case are comparable with experimental results. 

The spin wave function with different helicities of quark and antiquark can be expressed as
\eq 
\phi_\pi(\uparrow, \uparrow)&=& - \dfrac{k_{1} - i k_{2}}{x(1-x)},\nonumber\\
\phi_\pi(\uparrow, \downarrow)&=& \dfrac{m}{x(1-x)}+M_{\pi},\nonumber\\
\phi_\pi(\downarrow, \uparrow)&=& -\dfrac{m}{x(1-x)}-M_{\pi},\nonumber\\
 \phi_\pi(\downarrow, \downarrow)&=&- \dfrac{k_{1} + i k_{2}}{x(1-x)},
 \label{spin-part-wf}
\en
where $ M_{\pi} $ is the mass of pion.
Using the above expressions in Eq. \eqref{pion-state} and comparing it with Eq. \eqref{light-front-amplitude}, the LFWFs for pion state can be mapped out into different angular-momentum components as follows
\eq
\Psi_\pi^{(0)}(x,\bfk)&=& - \dfrac{m + M_{\pi}x(1-x)}{\sqrt{2}x(1-x)} \, \psi_\pi(x,\bfk),\nonumber\\
\Psi_\pi^{(1)}(x,\bfk)&=&  - \dfrac{1}{\sqrt{2}x(1-x)} \, \psi_\pi(x,\bfk).
\label{spin-improved-wf}
\en
\section{Transverse momentum dependent parton distribution functions (TMDs) of pion}

The quark transverse momentum distributions (TMDs) can be obtained by using quark-quark correlator for SIDIS  \cite{Barbara:2008, Meibner:2007, Boer:1998}
\be
 \Phi(x,\bfk)^{[\Gamma]} = \int\frac{dy^{-}d^{2}\textbf{\textit{y}}_{\perp}}{{2 \pi}^3}\,  e^{i\textit{k}.y} \langle \pi(P)\vert \bar{\psi}(0)\, {\cal W}[\,0\,\vert\infty, 0, 0_{T}]\, {\cal W}[\infty,y^{+}, y_{T} \vert\, y\,]\,\psi(\textit{y}) \vert \pi(P)\rangle \Big\vert_{{\textit{y}}^{+}=0},
\label{correlator}
\ee
where $ \Gamma $ is the Dirac operator, $ k $ is the momentum of the struck quark inside the pion of momentum $ P $ and $ x=p^{+}/P^{+} $ is the longitudinal momentum fraction carried by struck quark. Here $ {\cal W} $ is the  gauge link operator or Wilson line which makes the correlation function gauge-invariant. Wilson  line connects the two quark fields $ \psi $ at different points $0$ and $ \textit{y} $ and is defined as 
\eq
{\cal W}[0 \vert y] = {\cal P}\, exp\Big(-i g \int^{y}_{0} d\eta^{\mu} A_{\mu}(\eta)\Big),
\label{wilson-line}
\en
where $ A_{\mu} $ represents the gauge field. 

At leading twist, the transverse structure of pion is described by two TMDs: the unpolarized parton distribution function $f_{1 \pi}^{q}(x,\bfk^{2}) $ (also called T-even TMD) and the Boer-Mulders function $ h_{1 \pi}^{\bot}(x,\bfk^{2})$ (also called T-odd TMD). The leading twist pion TMDs can be  parametrized from the correlator given in  Eq. \eqref{correlator} and we have
\eq
 \dfrac{1}{2} \Phi^{[\gamma^{+}]} &=& f_{1 \pi}(x,\bfk^{2})
  ,\nonumber\\
 \dfrac{1}{2} \Phi^{[i \sigma^{j +} \gamma_{5}]} &=& \dfrac{\varepsilon^{i j} k_{\bot}^{j}}{M_{\pi}} h_{1 \pi}^{\bot}(x,\bfk^{2}).
 \label{pion-tmds}
 \en  
  
\subsection{Unpolarised parton distribution function of pion}
 The unpolarized  distribution  function $  f_{1 \pi}^{q}(x,\bfk^{2}) $ describes the probability of finding a quark with longitudinal momentum fraction $ x $ and transverse momentum $ k_{\bot} $ inside an unpolarized pion. One can obtain $  f_{1 \pi}^{q}(x,\bfk^{2}) $ from the lowest order  correlation  function \cite{Lu:2005} (without Wilson line) and we  have
 \eq  
f_{1 \pi}^{q}(x,\bfk^{2}) &=& \int\frac{dy^{-}d^{2}\textbf{\textit{y}}_{\perp}}{{2\pi}^3}\,  e^{i(y^{-}k^{+} -  \textbf{y}_{\bot}.\bfk)}\,\langle \pi(P)\vert \bar{\psi}(0)\,\gamma^{+} \,\psi(\textit{y}) \vert \pi(P)\rangle \Big\vert_{{\textit{y}}^{+}=0}.
\label{T-even-tmd}
\en
Using the light-front state of pion given in Eq. \eqref{light-front-amplitude} and the quark fields in terms of Fock operators, the unpolarized quark distribution $  f_{1 \pi}^{q}(x,\bfk^{2}) $ of pion can be expressed as the overlap of LFWFs given in Eq. \eqref{spin-improved-wf} as
\eq
f_{1 \pi}^{q}(x,\bfk^{2})&=& \dfrac{1}{(2 \pi)^3}\big[ \vert \psi^{(0)}_\pi(x,\bfk)\vert^{2}+\bfk^{2} \, \vert\psi^{(1)}_\pi(x,\bfk)\vert^{2}
\big].
\en
On solving analytically, we have the unpolarized parton distribution function of pion as
\eq
f_{1 \pi}^{q}(x,\bfk^{2})&=& \frac{N_{0}^{2}}{\pi }\, \frac{[\bfk^2+(m+x(1-x)M)^2]}{\kappa^2 \, x^3 \,(1-x)^3 }\, \exp\bigg[-\frac{\bfk^2 + m^2)}{\kappa^2 x (1-x)}\bigg].
\label{result-of-T-even-tmd}
\en
After integration over $ k_{\bot} $, the unpolarized  distribution  function  $  f_{1 \pi}^{q}(x,\bfk^{2}) $ reduces to parton distribution function  $f_{\pi}(x)$. The parton distribution function $f_{\pi}(x)$ satisfies the normalization condition
\eq
\int \mathrm{d} x f_{\pi}(x) = 1.
\label{normalization}
\en
It would be important to mention here that, in this model, the distribution  $ f_{1 \pi}^{q}(x,\bfk^{2}) $ of quark  with longitudinal momentum fraction $ x $, is equal to the distribution $ f_{1 \pi}^{\bar{q}}(1-x,\bfk^{2}) $ of an antiquark with longitudinal momentum fraction $ 1-x $ inside the pion.
 \begin{figure}
\includegraphics[width=2.5in]{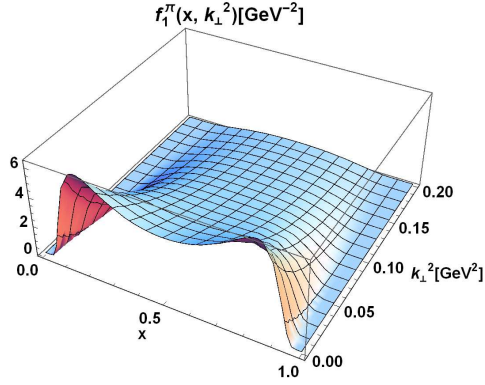}
\caption{3D plot of unpolarized TMD of pion$ f_{1 \pi}^{q}(x,\bfk^{2}) $  as a function of $ x $ and $ \bfk^2 $.}
\label{3d-pion-T-even-tmd}
\end{figure}

To obtain the numerical results of TMDs we need 3 parameters: the scale constant $\kappa$, mass of the quark $m$ and mass of the pion $ M_\pi $. We have  used the following set of parameters following Ref. \cite{Ahmady:2018}
\begin{equation}
 \kappa = 523  ~{\rm MeV}, ~~~~~ m = 330 ~{\rm MeV}~~~~~ {\rm and}~~~~~  M_\pi = 139~ {\rm MeV} .
\end{equation}
Using the above input parameters and to obtain the complete picture of distribution inside the pion, in Fig. \ref{3d-pion-T-even-tmd}, we present the 3D plot of unpolarized TMD of pion $ f_{1 \pi}^{q}(x,\bfk^{2}) $ as a function of $ x $ and $ \bfk^2 $. It is clear from the plot that at smaller values of $ \bfk^2 $ ($ < 0.1 $), the distribution has a double peak around $  x \approx 0.15 $ and $ x  \approx 0.8 $. The symmetric nature of unpolarized quark distribution $ x f_{1 \pi}^{q}(x, \bfk^2)$ of pion (given in Eq.   \eqref{spin-improved-wf}) under the exchange of $ x \longrightarrow 1-x $ is responsible for such a behavior. At larger values of $ \bfk^2 $, the two peaks merge and the distribution becomes Gaussian. On the other hand, at any fixed longitudinal momentum fraction $ x $,  the  amplitude of the distribution decreases with the increase in square of transverse momentum $ \bfk^2 $. 

\begin{figure}
\small{(a)}\includegraphics[width=2.5 in]{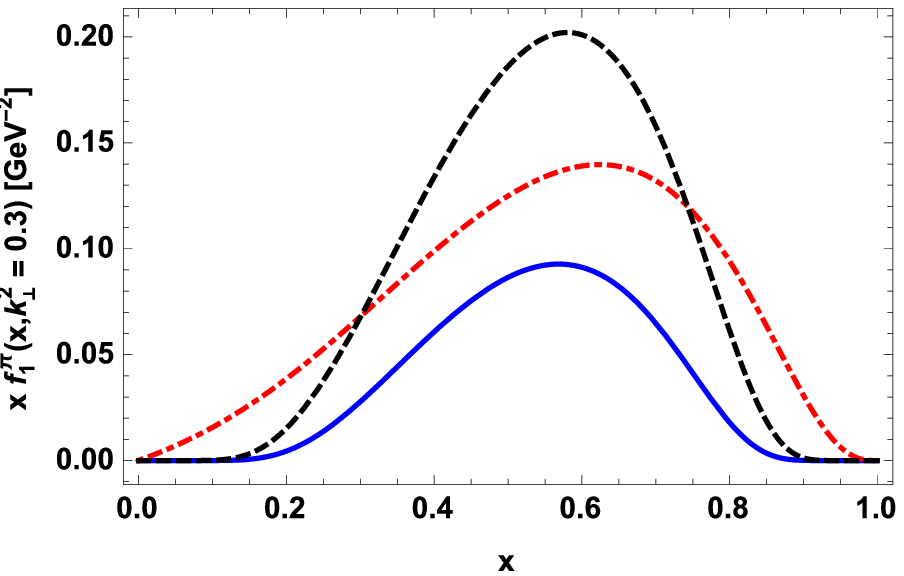}\hfill
\small{(b)}\includegraphics[width=2.5 in]{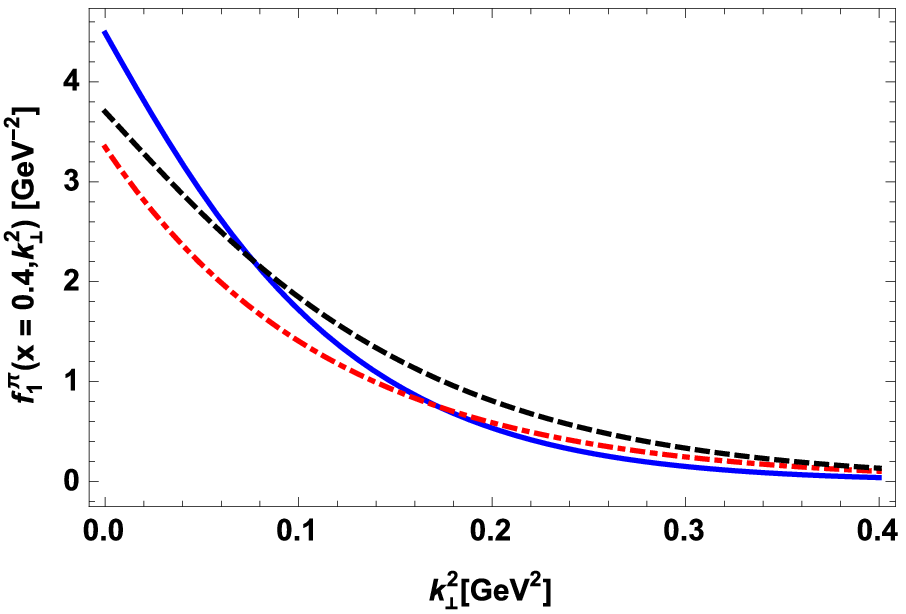}
\caption{Plots of unpolarized TMD $ x f_{1 \pi}^{q}(x, \bfk^2)$ of pion in our model (Solid line) (a) as a function of $x$ for a fixed value of $ \bfk^2 = 0.3 $ GeV$^2$ (b) as a function of $\bfk^2$ for a fixed values of $ x = 0.4 $. The plots also show the results of light-front constituent model \cite{Barbara:2014} (Dashed line) and Soft-wall AdS/QCD model given in \cite{Kaur:2018} (Dot-dashed line).}
\label{pion-T-even-tmd}
\end{figure}

The observations can be explained in detail from Fig. \ref{pion-T-even-tmd}  where we have shown the plot of $ x f_{1 \pi}^{q}(x, \bfk^2)$ as a function of $x$ for a fixed values of $ \bfk^2 = 0.3 $ Fig. \ref{pion-T-even-tmd} (a) and the  plot of unpolarized TMD of pion  $ f_{1 \pi}^{q}(x, \bfk^2)$  as a function of $\bfk^2$ for the fixed values of $ x = 0.4 $ in Fig. \ref{pion-T-even-tmd} (b). In both the cases we have also presented the available results from other models. We have compared our results with light-front constituent model (LFCM) \cite{Barbara:2014} and soft wall AdS/QCD model \cite{Kaur:2018}. In general, we observe that the overall behavior for the distribution function in the present model and other models  is similar.  The  amplitudes of different model results are however different. This may be due to different model assumptions and parameters used in the models. The peak of distribution in our model occurs around $ x = 0.6 $. The width of distribution peak is narrower in our model as compared to distribution in LFCM and soft wall AdS/QCD model. The width of the peak depends on the square of transverse momentum as discussed in Fig. \ref{3d-pion-T-even-tmd}.  To elaborate this, from Fig. \ref{pion-T-even-tmd} (b), we observe that at smaller values of $ \bfk^2 $, the amplitudes of the distribution in different models are different but as the value of $ \bfk^2 $ increases, the amplitude merges. This implies that when the transverse momentum carried by the quarks is large, the distribution is model independent.   

\subsection{Boer-Mulders function of the pion} 
Boer-Mulders function gives the distribution of transversely polarized partons inside an unpolarized pion. The Boer-Mulders function is obtained from the final-state interaction in SIDIS and initial-state interaction in DY-process via one gluon exchange. The sign of the Boer-Mulders function is however opposite for the both processes.
In SIDIS,  one can obtain the Boer-Mulders function $ h_{1 \pi}^{\bot}(x,\bfk^{2}) $ of the pion using the gauge link in  Eqs. \eqref{correlator} and \eqref{pion-tmds} as
\eq
h_{1 \pi}^{\bot}(x,\bfk^{2})_{\rm SIDIS} &=& \varepsilon^{i j} k_{\bot}^{j} \dfrac{M_{\pi}}{2 \bfk^2} \int\frac{dy^{-}d^{2}\textbf{\textit{y}}_{\perp}}{(2\pi)^3} e^{i(y^{-}k^{+} -  \textbf{y}_{\bot}.\bfk)}\nonumber\\
&& \langle \pi(P)\vert \bar{\psi}(0)\,{\cal W}[\,0\,\vert\infty, 0, 0_{T}]\,i \sigma^{i+}\gamma_{5} {\cal W}[\infty,y^{+}, y_{T} \vert\, y\,] \,\psi(\textit{y}) \vert \pi(P)\rangle,
\label{T-odd-tmd}
\en
where $ M_{\pi} $ is the mass of the pion. The gauge link is expanded upto the next-to-leading order to take into account the contribution from the one-gluon exchange between the struck quark and antiquark. Consequently,  one unit of orbital angular momentum transfer occurs which is compensated by the helicity flip of quark from the initial to final state of pion \cite{Brodsky:2002, Brodsky:2002(2), Barbara:2010, Barbara:2014}.

After solving the above correlator, we have obtained the following form of Boer-Mulders function $ h_{1 \pi}^{\bot}(x,\bfk^{2}) $ corresponding to SIDIS
\eq
h_{1 \pi}^{\bot}(x,\bfk^{2})_{\rm SIDIS} &=&\dfrac{4}{3} g^2 M_{\pi} \dfrac{k_{\bot}^{-}}{\bfk^2}\,\int \frac{d^2\bfq}{(2\pi)^5} \dfrac{1}{\bfq^2}\nonumber\\
& \times & \, [k_{\bot}^{+} \Psi_\pi^{(0)*}(x, \bfk) \Psi_\pi^{(1)}(x^{'},\bfk^{'})- k_{\bot}^{'+} \Psi_\pi^{(0)}(x,\bfk) \Psi_\pi^{(1)*}(x^{'},\bfk^{'})],
\en 
where $ k_{\bot}^{\pm} = k_{x} \pm i k_{y} $ and $ \bfk^{'}=\bfk - \bfq $. By inserting the LFWF from Eq. \eqref{spin-improved-wf} in the above equation, the following analytic expression for $ h_{1 \pi}^{\bot}(x,\bfk^{2}) $ can be derived
\eq
h_{1 \pi}^{\bot}(x,\bfk^{2})_{\rm SIDIS} &=&\dfrac{4}{3} g^2 M_{\pi} \dfrac{(4 \pi N_0)^2}{\bfk^2}\,\nonumber\\
& \times & \, \dfrac{(m + M_{\pi} x(1-x))^{2}}{2 x^2 (1-x)^2} \Big[ 1 - \exp{ \left[\frac{\bfk^2}{2 \kappa^2 x(1-x)} \right]} \Big] \exp{ \left[ -\frac{\bfk^2 + m^2}{\kappa^2 x(1-x)} \right]}.
\en
Here $ g $ is the parameter related to the coupling constant as $ g^2 = 4 \pi \alpha_{s}(\mu_{0}) $. We have obtained the result for Boer-Mulders function $ h_{1 \pi}^{\bot}(x,\bfk^{2}) $ of pion in DY-process by changing the sign i.e. $  h_{1 \pi}^{\bot}(x,\bfk^{2})_{\rm DY} $ = $ - h_{1 \pi}^{\bot}(x,\bfk^{2})_{\rm SIDIS} $. 
The model calculation is constrained  by the model-independent positivity relation of unpolarized TMD and Boer-Mulders function is defined as
\eq
 P^\pi(x,\bfk^2) \equiv f_{1 \pi}(x,\bfk^2) - \dfrac{\bfk}{M_{\pi}} \vert h_{1 \pi}^{\bot}(x,\bfk^{2}) \vert \geq 0.
 \en
 \begin{figure}
\includegraphics[width=2.5in]{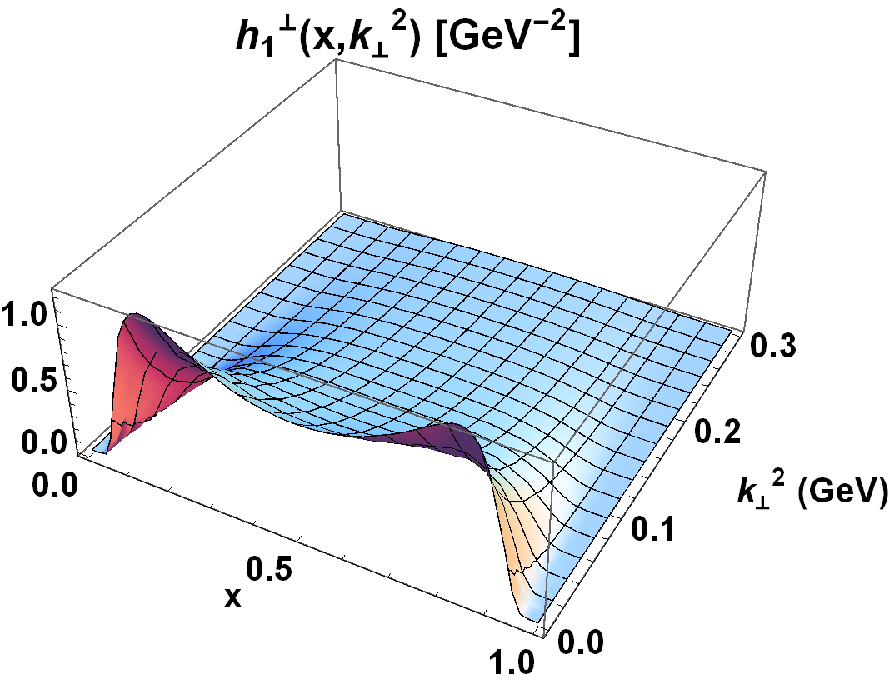}
\caption{3D plot of Boer-Mulders function $h_{1 \pi}^{\bot}(x,\bfk^{2})$ in DY-process as a function of $ x $ and $ \bfk^2 $.}
\label{3d-pion-tmd}
\end{figure}

We have presented the 3D plot of Boer-Mulders function in DY-process as a function of $ x $ and $ \bfk^2 $  in Fig. \ref{3d-pion-tmd}.  The amplitude of $  h_{1\pi}^{\bot}(x, \bfk^2)$ decreases with the  increase in $\bfk^2$ for any fixed value of $ x $. This implies that the probability of finding the quarks with transverse spin inside the pion is less when they carry large transverse momentum. At smaller values of $\bfk^2$, a double peak is observed similar to the case of  unpolarized quark distribution which is because of the symmetric nature of Boer-Mulders function under the exchange of $ x \longrightarrow 1-x $. Our results are in agreement with the results for pion Boer-Mulders function in light-front constituent approach \cite{Barbara:2014}.

\begin{figure}
\small{(a)}\includegraphics[width=2.5 in]{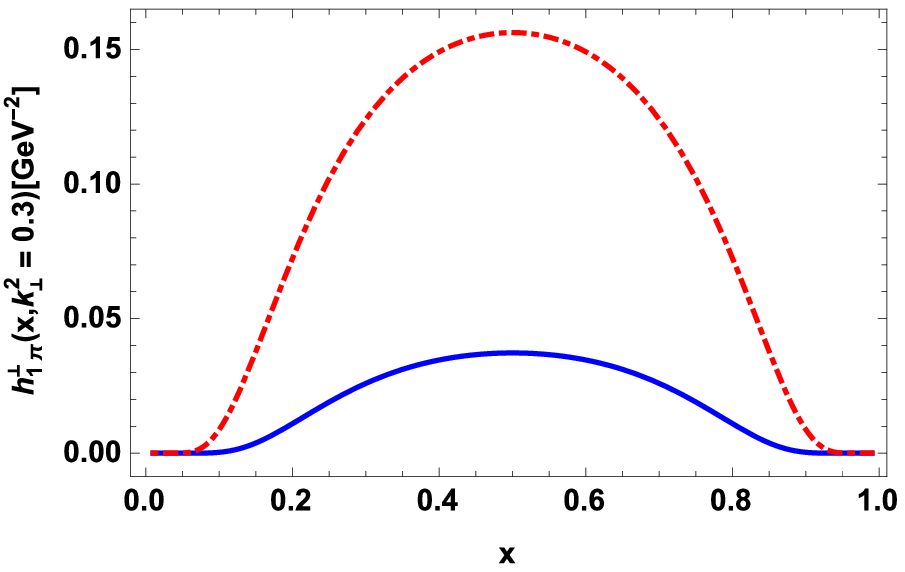}\hfill
\small{(b)}\includegraphics[width=2.5 in]{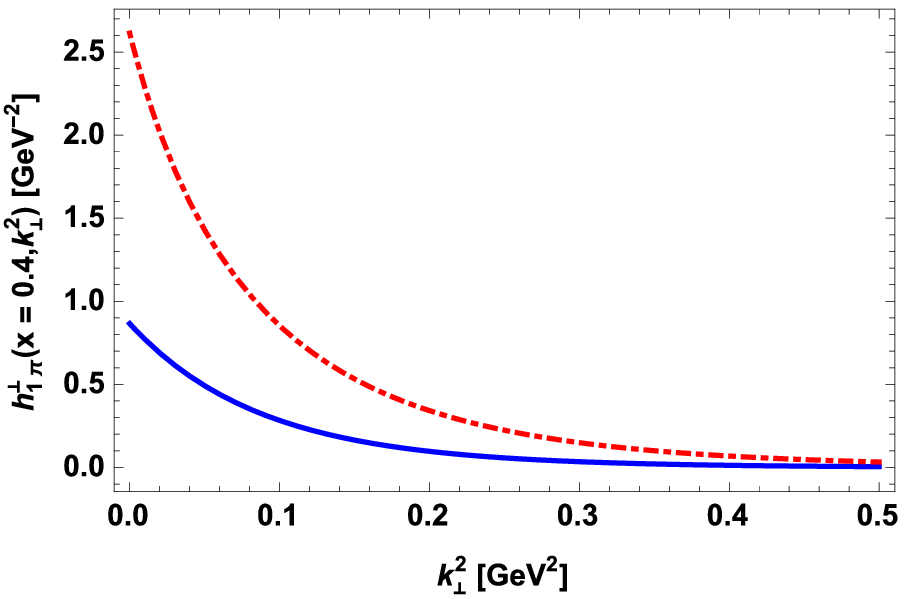}
\caption{Plot of Boer-Mulders function of pion $  h_{1\pi}^{\bot}(x, \bfk^2)$ in DY-process in our model (Solid line) (a) as a function of $x$ for a fixed value of $ \bfk^2 = 0.3 $ GeV$^2$ (b) as a function of $\bfk^2$ for a fixed values of $ x = 0.4 $. We compare our results with Light-front constituent model (Dot-dashed line).}
\label{pion-BM}
\end{figure}

In Fig. \ref{pion-BM} (a), we have presented  the plot of $  h_{1\pi}^{\bot}(x, \bfk^2)$  as a function of $x$ for a fixed value of $ \bfk^2 = 0.3 $ GeV$^2$ and have compared it with the result from light-front constituent model (LFCM) \cite{Barbara:2014}. The amplitude of Boer-Mulders function is different in two models but shape of the distribution is similar. We observe that the distribution is not purely Gaussian in both the models.     
We shown  the plot of $  h_{1\pi}^{\bot}(x, \bfk^2)$ as a function of $\bfk^2$ for a fixed value of $ x = 0.4 $ and compare it with LFCM in Fig. \ref{pion-BM} (b). The distributions in both the models have similar behavior at larger value of $\bfk^2$ but at smaller value of $\bfk^2$, the amplitudes are different. In $0 \leq \bfk^2 \leq 0.1 $ region, the distribution in LFCM falls off rapidly with $\bfk^2$ as compared to the case in our model.

\begin{figure}
\includegraphics[width=2.5in]{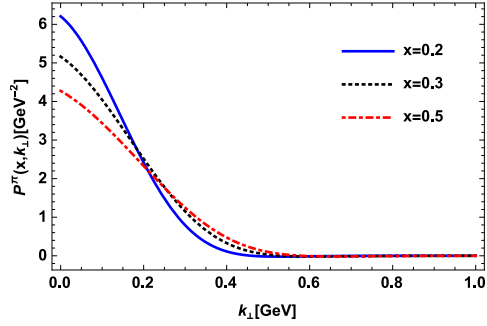}
\caption{The positivity relation for the quark in pion as a function of $\bfk $ for the different values of $ x $.}
\label{positivity}
\end{figure}

The model results for positivity at different values of $ x $ are plotted in Fig. \ref{positivity}. It is observed that the inequality is very well satisfied in this model even at the large value of transverse momentum $ \bfk $. It can be concluded from here that the amplitude of unpolarized TMD is always greater than the amplitude of Boer-Mulders function. This implies that in this model, the probability of finding an unpolarized quark inside the pion is larger than the probability of the finding transversely polarized quark in the pion.

\section{Evolution of unpolarized TMD}
The result of unpolarized TMD obtained in Eq. \eqref{result-of-T-even-tmd} is at a scale of 0.316 GeV. However, to compare the present results with the future experimental data, the unpolarised TMD needs to be evolved to higher scales. The evolution of TMD in the coordinate space ($ b_{\perp} $-space) is comparatively  more convenient than that in the momentum space ($ \bfk $-space). 
The pion TMD evolution from initial scale to another scale in the coordinate space ($ b_{\perp} $) can be obtained using \cite{Aybat:2011, Collins, Rogers, Collins:2015}
\eq
\tilde{f}_{1 \pi}(x,b_{\bot}; \mu)= \exp^{\tilde{S}(\bar{b_\ast};\mu_b,\mu)}\, \exp^{g_K(b_\bot) \ln\frac{\mu}{\mu_0}}\,\tilde{f}_{1 \pi}(x,b_{\bot}; \mu_0).
\label{evolution}
\en
Here, the Sudakov form factor $\tilde{S}(\bar{b_\ast};\mu_b,\mu)$ is spin independent and is considered upto next-to-leading logarithmic (NLL) approximation and at leading order in $ \alpha_{s} $  \cite{Frixione, Echevarria}. The nonperturbative function $ g_{K}(b_{\bot}) $ in Eq. \eqref{evolution} is parametrized in quadratic form  as $ g_{K} = - g_2 b_{\bot}^2 / 2 $ \cite{Aybat:2012, Anselmino:2012, Landry}.

In the small $b$ region, the $b-$dependent TMD is perturbatively calculable but in the large $b$ region, the dependence turns out to be nonperturbative. To obtain a complete information of perturbative and nonperturbative part, a parameter $b_{max}$ can be introduced and we have
\eq
\mu_b=\frac{C_1}{b_*(b_\bot)}, \quad \quad b_*(b_\bot)=\frac{b_\bot}{\sqrt{1+\frac{b^2_\bot}{b^2_{max}}}}.
\en
In the above expression, the constant $ C_1 $ adopts a particular choice $ C_1=2 e^{-\gamma_E} $ \cite{Aybat:2011, Aybat:2012}, where $\gamma_E=0.577$ is the Euler constant \cite{Collins:1985}. A $b-$dependent function $ b^\ast $ is defined to have the property $ b^{\ast} \approx b $ at low values of $b$ and $ b^{\ast} \approx b_{max} $ at large values of $ b $. The typical value of $ b_{max} $ has been chosen such that $ b^\ast $ is always in the perturbative region. Therefore,  $ b_{max} $ is anticorrelated to the nonperturbative parameter $ g_2 $. From the experimental data, $ g_{2} $ parameter can be extracted with a fixed value all other parameters. We have taken two values of $ g_{2} = 0.09, 0.13 $ following \cite{Bacchetta:2017} for the present work.

\begin{figure}
\includegraphics[width=2.5in]{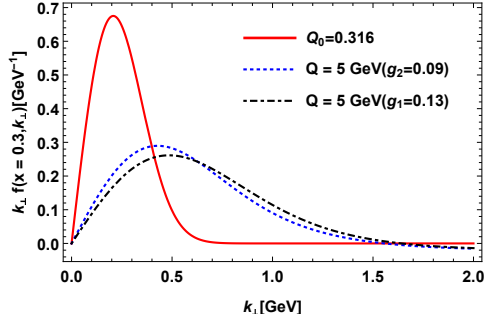}
\caption{Unpolarized TMD as a function of  $ \bfk^2 $ for a fixed value of $x$ at $0.3$  evoluted from initial scale $ Q_{0} = 0.316 $ upto the scale $ Q = 1 $ GeV with two different values of $ g_2 $.}
\label{3d-pion-tmd-evoluted}
\end{figure}
In Fig. \ref{3d-pion-tmd-evoluted}, we have presented the results of unpolarized TMD as a function of $ \bfk^2 $ for a fixed value of $x$ at $0.3$  after evolution upto the scale $ Q = 1 $ GeV. We have performed the evolution for two different values of $ g $. It can be clearly seen that, after evolution, the peak of the distribution shifts from $ \bfk = 0.2$ GeV to $ \bfk = 0.4$ GeV for $ g_2 = 0.09 $ and to $ \bfk = 0.5 $ for $ g_2 = 0.13 $. Also it is observed that the width of the unpolarized TMD at initial scale $ Q_{0} = 0.316 $ GeV  broadens to the scale $ Q = 1 $ GeV after evolution. The broadening of distribution further depends on the value of $ g_2 $. The possible explanation for the broadening of the peak can be the nonperturbative component of TMD evolution which introduces a component of gluon radiation and is responsible for the broadening of distribution. 
\section{Summary and conclusions}
In this paper, we have studied the transverse momentum dependent parton distribution functions of pion at leading twist. We have obtained the unpolarized parton distribution function $ f_{1 \pi}(x, \bfk) $ and Boer-Mulders function $ h_{1 \pi}^{\bot}(x,\bfk^{2}) $ of pion by using the light-front holographic model with spin-improved wave functions. The result of $ f_{1 \pi}(x, \bfk) $ has been  compared with light-front constituent model and soft-wall AdS/QCD Model. The overall nature of distribution is similar in all three models with little difference in their amplitude.  The  amplitude of the distribution decreases with the increase in square of transverse momentum $ \bfk^2 $ for a fixed longitudinal momentum fraction $ x $. The symmetric nature of unpolarized quark distribution $ x f_{1 \pi}^{q}(x, \bfk^2)$ of pion under the exchange of $ x \longrightarrow 1-x $ leads to a double peak of the distribution. 
The width of distribution peak is narrower in our model as compared to distribution in LFCM and soft wall AdS/QCD model for a fixed transverse momentum $ \bfk^2 $ which show that the probability of finding the partons is more in this region. Further, it is observed that the distribution is model independent when the transverse momentum carried by the quarks is large.

The results for $ h_{1 \pi}^{\bot}(x,\bfk^{2}) $ have been calculated from the correlator function with Wilson line.
It is found that the probability of finding the quarks with transverse spin inside the pion is less when they carry large transverse momentum. At smaller values of $\bfk^2$, a double peak is observed similar to the case of  unpolarized quark distribution.
The results are compared with the light-front constituent model. 
At  smaller values of transverse momentum, the amplitudes of distribution in both models  are different but at larger values of transverse momentum results they agree quite well. In $0 \leq \bfk^2 \leq 0.1 $ region, the distribution in LFCM falls off rapidly with $\bfk^2$ as compared to the case in our model.

From the results for positivity at different values of $ x $ it is observed that the amplitude of $ h_{1 \pi}^{\bot}(x,\bfk^{2}) $ is smaller than the $ f_{1 \pi}(x, \bfk) $ and the inequality is very well satisfied even at a large value of transverse momentum $ \bfk $. The amplitude of unpolarized TMD is always greater than the amplitude of Boer-Mulders function implying that the probability of finding an unpolarized quark inside the pion is larger than the probability of the finding transversely polarized quark in the pion.
We have also performed the evolution of $ f_{1 \pi}(x, \bfk) $ from initial model scale to higher scale. We have observed a width  broadening of distribution for evolution from initial scale $ Q_{0} = 0.316 $ GeV  to the scale $ Q = 1 $ GeV. The broadening of distribution depends on the value of nonperturbative parameter $ g_2 $ which introduces a component of gluon radiation. 

\acknowledgements
H.D. would like to thank the Department of Science and Technology (Ref No. EMR/2017/001549) Government of India for financial support. 



\begin{thebibliography}{99}
\bibitem{burk2000} M. Burkardt, Phys. Rev. D {\bf 62}, 071503 (2000); M. Burkardt, Phys. Rev. D {\bf 66}, 119903 (2002).
\bibitem{Ji:1997} X. Ji, Phys. Rev. D \textbf{55}, 7114 (1997).
\bibitem{Ji2:1997} X. Ji, Phys. Rev. Lett. \textbf{78}, 610 (1997).
\bibitem{Radyushkin:1997} A. V. Radyushkin, Phys. Rev. D \textbf{56}, 5524 (1997).
\bibitem{Ji:1998} X. Ji and J. Osborne, Phys, Rev. D \textbf{58}, 094018 (1998).
\bibitem{Ji2:1998} X. Ji, J. Phys. G: Nucl. Part Phys. \textbf{24}, 1181 (1998).
\bibitem{Blmlein:2000} J. Blumlein, B. Geyer and D. Robaschik, Nucl. Phys. B \textbf{560}, 283 (1999).
\bibitem{Goeke:2001} K. Goeke, M. V. Polyakov, M. Vanderhaeghen, Progr. Part. Nucl. Phys. \textbf{47}, 401 (2001).
\bibitem{Diehl:2003} M. Diehl, Phys. Rep. \textbf{388}, 41 (2003).
%
\bibitem{Tangerman95} R. D. Tangerman and P. J. Mulders, Phys. Rev. D \textbf{51}, 3357 (1995).
\bibitem{Kotzinian95} A. Kotzinian, Nucl. Phys. B \textbf{441}, 234 (1995).
\bibitem{Mulders97} P. J. Mulders and R. D. Tangerman, Nucl. Phys. B \textbf{461}, 197 (1996); P. J. Mulders and R. D. Tangerman, Nucl. Phys.  B \textbf{484}, 538 (1997).
\bibitem{Mulders98} D. Boer and P. J. Mulders, Phys. Rev. D \textbf{57}, 5780 (1998).
\bibitem{Mulders07} A. Bacchetta, M. Diehl, K. Goeke, A. Metz, P. J. Mulders and M. Schlegel, JHEP \textbf{0702}, 093 (2007).
\bibitem{Ji:2005} X. Ji, J. P. Ma and F. Yuan Phys. Rev. D \textbf{71}, 034005 (2005).
\bibitem{Drell:1970}  S. D. Drell and Tung-Mow Yan, Phys. Rev. Lett. {\bf 25}, 316 (1970); S. D. Drell and Tung-Mow Yan, Phys. Rev. Lett. {\bf 25},902(1970).
\bibitem{Christenson:1970} J. H. Christenson {\it et al.}, Phys. Rev. Lett. {\bf 25}, 1523 (1970).
\bibitem{Collins:2002} J. C. Collins, Phys. Lett. B \textbf{536}, 43 (2002).
\bibitem{Quintans:2011} C. Quintans  and the COMPASS Collaboration, J. of Phys. Conf. Ser. \textbf{295}, 012163 (2011). 
\bibitem{Baranov:2014} S. P. Baranov, A. V. Lipatov and N. P. Zotov, Phys. Rev. D \textbf{89}, 094025 (2014).
\bibitem{Boer:1998} D. Boer and P.  J. Mulders, Phys. Rev. \textbf{D 57}, 5780 (1998).
%
\bibitem{Brodsky:2002} S. J. Brodsky, D. S. Hwang and I. Schmidt, Phys. Lett. B \textbf{530}, 99 (2002).
%
\bibitem{Ji:2002} X. D. Ji and F. Yuan, Phys. Lett. B \textbf{543}, 66 (2002).
%
\bibitem{Brodsky:2002(2)} S. J. Brodsky, D. S. Hwang and I. Schmidt, Nucl. Phys. B \textbf{642}, 344 (2002).
%
\bibitem{Boer:2003} D. Boer, S. J. Brodsky and D. S. Hwang, Phys. Rev. D \textbf{67}, 054003 (2003).
%
\bibitem{Boer:2003(2)} D. Boer, P. J. Mulders and F. Pijlman, Nucl. Phys. B \textbf{667}, 201 (2003).
%
\bibitem{Belitsky:2003} A. V. Belitsky, X. Ji and F. Yuan, Nucl. Phys. B \textbf{656}, 165 (2003).
%
%
\bibitem{McGaughey:1999} P. L. McGaughey, J. M. Moss and J. C. Peng, Ann. Rev. Nucl. Part. Sci. {\bf 49}, 217 (1999).
%
\bibitem{Reimer:2007} P. E. Reimer, J. Phys. G {\bf 34}, S107 (2007).
%
\bibitem{Chang:2013} W. C. Chang and D. Dutta, Int. J. Mod. Phys. E {\bf 22}, 1330020 (2013).
%
\bibitem{Peng:2014} J. C. Peng and J.-W. Qiu, Prog. Part. Nucl. Phys. {\bf 76}, 43 (2014).
%
\bibitem{Altarelli:1996}  G. Altarelli, S. Petrarca and F. Rapuano, Phys. Lett. B \textbf{373}, 200 (1996).
%
\bibitem{Chang:2014} L. Chang, C. Mezrag, H. Moutarde, C. D. Roberts, J. R.-Quinterod and P. C. Tandy, Phys. Lett. B \textbf{737}, 23 (2014).
%
\bibitem{Shi:2018} C. Shi, C. Mezrag and  H.-S. Zong, Phys. Rev. D \textbf{98}, 054029 (2018)
%
\bibitem{Nguyen:2011} T. Nguyen, A. Bashir, C. D. Roberts and P. C. Tandy, Phys. Rev. C \textbf{83}, 062201 (2011).
%
\bibitem{Chang:2015} L. Chang and A. W. Thomas, Phys. Lett. B, \textbf{749}, 547-550 (2015).
\bibitem{Polyakov:1999} M. V. Polyakov and C. Weiss, Phys. Rev. D {\bf 60}, 114017 (1999).
\bibitem{Dalley:2001} S. Dalley, Phys. Rev. D {\bf 64}, 036006 (2001).
\bibitem{Davidson:2002} R. M. Davidson and E. R. Arriola, Acta. Phys. Polon. B {\bf 33}, 1791 (2002).
\bibitem{Dalley:2003} S. Dalley and B. V. Sande, Phys. Rev. D {\bf 67}, 114507 (2003).
%
\bibitem{Broniowski:2003} W. Broniowski and E. R. Arriola, Phys. Lett. B {\bf 574}, 57 (2003).
%
\bibitem{Bromel:2008}  D. Bromel {\it et al.}, Phys. Rev. Lett. {\bf 101}, 122001 (2008).
%
\bibitem{Frederico:2009} T. Frederico, E. Pace, B. Pasquini and G. Salme, Phys. Rev  D {\bf 80}, 054021 (2009).
\bibitem{Dorokhov:2011} A. E. Dorokhov, W. Broniowski and E. R. Arriola, Phys. Rev. D {\bf 84}, 074015 (2011).
\bibitem{Mezrag:2014} C. Mezrag, L. Chang, H. Moutarde, C. D. Roberts, J. R.-Quintero , F. Sabatié and S. M. Schmidt, Phys. Lett. B \textbf{741}, 190 (2015).
%
\bibitem{Thomas:2015} T. Gutsche, V. E. Lyubovitskij,
I. Schmidt and A. Vega, J. Phys. G: Nucl. Part. Phys. \textbf{42}, 095005 (2015).
%
\bibitem{Barbara:2014} B. Pasquini and P. Schweitzer, Phys. Rev. D {\bf 90}, 014050 (2014).
%
\bibitem{Lorce:2016} C. Lorcé, B. Pasquini and P. Schweitzer, Eur. Phys. J.  C \textbf{76}, 7, 415 (2016). 
%
\bibitem{Lu:2004} Z. Lu and B. Q. Ma, Nucl. Phys. A \textbf{741}, 200 (2004). 
%
\bibitem{Meissner:2008} S. Meissner, A. Metz, M. Schlegel and K. Goeke, J. High Energy Phys. \textbf{08}, 038 (2008). 
%
\bibitem{Gamberg:2010} L. Gamberg and M. Schlegel, Phys. Lett. B \textbf{685}, 95 (2010)
%
\bibitem{Lu:2004(2)}  Z. Lua, B.-Q. Ma, Phys. Rev. D \textbf{70}, 094044 (2004). 
%
\bibitem{Noguera} S. Noguera and S. Scopetta, J. High Energy Phys. , 102 (2015)
%
\bibitem{Bacchetta:2017} A. Bacchetta, S. Cotogno and B. Pasquini, Phys. Lett. B \textbf{771}, 546-552 (2017).
%
\bibitem{Kaur:2018} N. Kaur, N. Kumar, C. Mondal and H. Dahiya, Nucl. Phys. B \textbf{934}, 80-95 (2018).
%
\bibitem{Brodsky2005} G. F. de Teramond and S. J. Brodsky, Phys. Rev. Lett. \textbf{94}, 201601 (2005).
%
\bibitem{Brodsky2006} S. J. Brodsky and G. F. de Teramond, Phys. Rev. Lett. \textbf{96}, 201601 (2006).
%
\bibitem{Brodsky2009} G. F. de Teramond and S. J. Brodsky, Phys. Rev. Lett. \textbf{102}, 081601 (2009).
%
\bibitem{Brodsky2014} S. J. Brodsky, G. F. de Teramond and H. G. Dosch, Phys. Lett. B \textbf{729}, 3(2014).
%
\bibitem{Brodsky15} S. J. Brodsky, G. F. de Teramond, H. G. Dosch and J. Erlich, Phys. Rept \textbf{584}, 1 (2015).
%
\bibitem{Brodsky:2009} S. J. Brodsky and G. F. de Teramond, Subnucl. Ser. \textbf{45}, 139 (2009).
%
\bibitem{Vega:2009} A. Vega and I. Schmidt, Phys. Rev. D \textbf{79}, 055003 (2009).
%
\bibitem{Vega:2009(2)} A. Vega, I. Schmidt, T. Branz, T. Gutsche and V. E. Lyubovitskij, Phys. Rev. D \textbf{80}, 055014 (2009).
%
\bibitem{Rohit:2015} R. Swarnkar and D. Chakrabarti, Phys. Rev. D \textbf{92}, 074023 (2015).
%
\bibitem{Ahmady:2017} M. Ahmady, F. Chishtie and R. Sandapen, Phys. Rev. D \textbf{95}, 074008, (2017).
%
\bibitem{Ahmady:2018} M. Ahmady, C. Mondal and R. Sandapen, Phys. Rev. D \textbf{98}, 034010 (2018).
%
\bibitem{Barbara:2010} B. Pasquini and F. Yuan, Phys. Rev. D {\bf 81}, 114013 (2010).
\bibitem {Barbara:2008} B. Pasquini and P. Schweitzer, Phys. Rev. D \textbf{78}, 071502 (2008).
\bibitem {Meibner:2007} S. Meissner, A. Metz and K. Goeke, Phys. Rev.\textbf{ D 76}, 034002 (2007).
\bibitem{Lu:2005} Z. Lua and B.-Q. Ma, Phys. Lett. B \textbf{615}, 200–206 (2005).
\bibitem{Aybat:2011} S. M. Aybat and T.~C.~Rogers,
  Phys.\ Rev.\ D {\bf 83}, 114042 (2011).
\bibitem{Collins} J. Collins, Cambridge Monographs on Particle Physics, Nuclear Physics and Cosmology, \textbf{32} Cambridge University Press (2011).
\bibitem{Rogers} T. C. Rogers, Eur. Phys. J. A \textbf{52} (6), 153 (2016).
\bibitem{Collins:2015} J. Collins and T. Rogers, Phys. Rev. D \textbf{91}, 074020, (2015).
\bibitem{Frixione} S. Frixione, P. Nason and G. Ridolfi, Nucl. Phys. B \textbf{542}, 311 (1999).
\bibitem{Echevarria} M. G. Echevarria, A. Idilbi, A. Schäfer and I. Scimemi, Eur. Phys. J. C \textbf{73}(12), 2636 (2013).
\bibitem{Anselmino:2012} M. Anselmino, M. Boglione and S. Melis,
  Phys.\ Rev.\ D {\bf 86}, 014028 (2012).
\bibitem{Aybat:2012} S. M. Aybat, J. C. Collins, J. W. Qiu and T. C. Rogers, 
 Phys. Rev. D {\bf 85}, 034043 (2012).
\bibitem{Landry} F. Landry, R. Brock, P. M. Nadolsky and C. P. Yuan, Phys. Rev. D \textbf{67}, 073016 (2003).
\bibitem{Collins:1985} J. C. Collins, D. E. Soper and G. F. Sterman,
  Nucl.\ Phys.\ B {\bf 250}, 199 (1985).
%
%
\end{thebibliography}
\end{document}